\documentclass[npg,manuscript]{copernicus-arxiv}
\usepackage[section]{placeins}
\begin{document}
\nolinenumbers
\renewcommand{\acknowledgementsname}{Acknowledgments}
\title{When is logistic error growth a spectral reduction?}
\Author[1][mpena@uconn.edu]{Malaquias}{Pe\~na}
\affil[1]{Eversource Energy Center, University of Connecticut, Storrs, CT 06269, USA}
\runningtitle{Interpreting logistic forecast-error growth}
\runningauthor{M. Pe\~na}
\received{}\pubdiscuss{}\revised{}\accepted{}\published{}
\firstpage{1}
\maketitle
\renewcommand{\baselinestretch}{1}\normalsize
\begin{abstract}
Forecast-error growth has been studied through complementary approaches: scalar models describe the evolution of bulk error, whereas spectral theories follow decorrelation and error transfer across scales. Their connection matters when a fitted growth curve is used to infer a physical mechanism or a limit of predictability. We establish a conditional link between these approaches. Within a bilinear spectral exchange model, assume a stationary reference spectrum and a remainder that vanishes on the proposed common-shape evolution. For a common initial decorrelation fraction strictly between zero and one, the spectral shape is preserved and its amplitude grows logistically exactly when every participating scale shares one positive growth coefficient. The converse inference fails: a logistic aggregate need not imply proportional decorrelation or a common coefficient. An exact two-shell construction demonstrates this failure with identical aggregate curves and different spectral dynamics. For heterogeneous amplitudes, an exact averaging identity separates the contributions of growth-rate mismatch, shape variance, rate--shape covariance, evolving spectral weights, and residual tendencies. A finite-time bound relates their combined effect to departure from a reference logistic curve. Forced surface-quasigeostrophic simulations illustrate how similar retrospective agreement with a logistic reference can coexist with heterogeneous effective shell rates and substantial instantaneous tendency discrepancies. Thus a scalar growth curve alone cannot identify how uncertainty is distributed across scales or justify interpreting a fitted rate as a common spectral growth coefficient. The framework specifies the additional spectral and budget evidence needed to give scalar predictability models that interpretation.
\end{abstract}

\introduction
\label{sec:introduction}
Forecast uncertainty develops across interacting scales. Small errors can initially grow approximately exponentially, whereas finite errors encounter nonlinear interactions, dissipation, and eventual decorrelation. A scalar growth curve provides an economical description of this evolution, but its economy raises an interpretive question: which properties of the underlying flow survive the reduction? This question matters when a fitted error-growth rate or saturation level is used to characterize predictability, compare forecast systems, or construct a compact representation of an ensemble.

Two complementary lines of research motivate the present question. The scalar line describes a bulk error measure with a low-dimensional growth law. It includes the short-time error model of \citet{Leith1978}, the saturating model of \citet{Lorenz1982}, the treatment of growth and model error by \citet{DalcherKalnay1987}, and the stochastic descriptions of \citet{Nicolis1992} and \citet{Bach2025}. These models use different error variables and assumptions: Leith's short-time law does not saturate, Lorenz's logistic law concerns an error norm, and later formulations include model-error sources or stochastic terms. They do not all prescribe the same logistic law for error energy.

The spectral line addresses how decorrelation develops and propagates across interacting scales. Building on multiscale predictability arguments \citep{Lorenz1969}, \citet{Leith1971} and \citet{LeithKraichnan1972} studied spectral error dynamics in turbulence; \citet{Rotunno2008} examined how spectral slope influences multiscale predictability. Subsequent experiments demonstrate the roles of fast and slow dynamics, scale-dependent saturation, and perturbation size \citep{Herrera2010,UboldiTrevisan2015}. These lines are not mutually exclusive: \citet{ZagarEtAl2017} fitted growth curves separately to zonal wavenumbers in an operational ensemble. Our question is when a scalar growth law represents a preserved spectral structure, rather than only a useful description of its aggregate.

A commonly used scalar description is the logistic equation for half-energy error $\varepsilon$,
\begin{equation}
 \dot\varepsilon=\lambda_0\varepsilon(1-\varepsilon/L),
 \qquad \lambda_0>0,\quad L>0.
 \label{eq:logistic-reference}
\end{equation}
Here $\varepsilon$ is one-half the squared error norm, $\lambda_0$ is its small-error growth rate, and $L$ is a fixed asymptotic error energy. Interpreting $L$ as the error energy of decorrelated states requires the reference-spectrum assumptions in Sect.~\ref{sec:ray}. Logistic and related models have long been useful in operational predictability studies \citep{DalcherKalnay1987,PenaToth2014,Zhang2019}. More recent work analyzes how unresolved scales and model error alter fitted growth parameters \citep{BednarKantz2024}. \citet{Bach2025} extend scalar error modeling with multiplicative noise, obtain agreement with mean and probabilistic features of operational forecast errors, and compare low- and high-wavenumber ranges. These studies establish useful scalar descriptions at different scales. Here we ask when aggregate logistic behavior represents a preserved spectral shape, and how to assess departures from that particular reduction.

The paper makes three connected contributions. First, it states a conditional invariant-ray proposition: within an assumed spectral exchange model, all participating scales retain the same decorrelation amplitude and that amplitude grows logistically. An exact two-shell example then demonstrates that a logistic aggregate does not imply this spectral behavior. Second, an averaging identity organizes departures from the common-amplitude model, including changes in spectral weights and terms outside the assumed local growth law. Applying the standard Gronwall inequality to the complete departure gives a sufficient finite-time error bound. The identities and inequality use standard mathematical tools; their purpose here is to specify what a spectral interpretation of logistic growth must account for.

Third, a forced surface-quasigeostrophic (SQG) ensemble illustrates what those conditions can and cannot establish from archived data. Its tangent-linear growth rate is measured independently of the finite-error curves. We distinguish diagnostics specified before their evaluation, later exploratory decompositions, and a subsequent audit of the stored physical budgets. These analyses reproduce the primary-band summaries but do not identify a closed spectral exchange operator or provide a continuous-trajectory approximation certificate. A separate covariance-coupling null test supplies an additional, unresolved test of physical interpretation. An alternative moving-front reduction is developed in Appendix~\ref{app:front}; its empirical use requires a separately established similarity interval.
\section{Exact budgets and conditional scalar reductions}
\label{sec:theory}
\subsection{Error energy, projection, and normalization}
\label{sec:exact-budget}
Let $u(t)$ be the reference state and $\hat u(t)$ its forecast in a real Hilbert space with inner product $\langle\cdot,\cdot\rangle$ and norm $\|\cdot\|$. The reference obeys $\dot u=\mathcal L u+\mathcal Q(u,u)+f$, where $f$ is the imposed forcing, $\mathcal L$ is linear, $\mathcal Q$ is bilinear, and $\langle v,\mathcal Q(v,v)\rangle=0$ for the chosen conservative norm. Perfect-model twins driven by the same forcing have error $\delta=\hat u-u$ and half-energy error $\varepsilon=\|\delta\|^2/2$. Subtraction and the conservative identity give
\begin{equation}
 \dot\varepsilon=\langle\delta,\mathcal L\delta+\mathcal Q(u,\delta)+\mathcal Q(\delta,u)\rangle.
 \label{eq:exact-error-budget}
\end{equation}
The error self-interaction cancels in the full-domain inner product. Equation~(\ref{eq:exact-error-budget}) remains exact at finite amplitude: the error state changes nonlinearly even though the instantaneous energy production has this linearized-operator form. The finite-error trajectory still follows the full nonlinear error equation.

A spectral band is a fixed set of modes, and $\Pi_b$ denotes the orthogonal projection onto band $b$. For the SQG application, $\Pi_b$ commutes with the linear damping operator. Let $\delta_b=\Pi_b\delta$ and $\varepsilon_b=\|\delta_b\|^2/2$. The exact projected balance is
\begin{equation}
 \dot\varepsilon_b=P_{\mathrm{linear}}+P_{\mathrm{TL,nl}}+P_{\mathrm{pure}},\qquad
 P_{\mathrm{pure}}=\langle\delta_b,\Pi_b\mathcal Q(\delta,\delta)\rangle.
 \label{eq:band-budget}
\end{equation}
Here $P_{\mathrm{linear}}=\langle\delta_b,\Pi_b\mathcal L\delta\rangle$ and
$P_{\mathrm{TL,nl}}=\langle\delta_b,\Pi_b[\mathcal Q(u,\delta)+\mathcal Q(\delta,u)]\rangle$.
The label TL denotes the tangent-linear operator, obtained by linearizing the dynamics about the reference state; its nonlinear contribution is evaluated here on the finite error. The last term represents conservative interband redistribution and can take either sign. A numerical evaluation must also retain its temporal-discretization residual. Nonidentical forcing would add an error-forcing term; it is absent for the common-forcing twins considered here.

Write $u_b=\Pi_bu$ and $\hat u_b=\Pi_b\hat u$. With $E_{t,b}=\|u_b\|^2/2$, $E_{f,b}=\|\hat u_b\|^2/2$, and $C_b=\langle u_b,\hat u_b\rangle/2$, define $S_b=E_{t,b}+E_{f,b}>0$. Then
\begin{equation}
 \varepsilon_b=S_b-2C_b,\quad x_C=\frac{2C_b}{S_b},\quad a_b=1-x_C=\frac{\varepsilon_b}{S_b}.
 \label{eq:normalization}
\end{equation}
The normalized tendency is consequently
\begin{equation}
 \dot a_b=\frac{P_{\mathrm{linear}}+P_{\mathrm{TL,nl}}+P_{\mathrm{pure}}}{S_b}
          -a_b\frac{\dot S_b}{S_b}.
 \label{eq:normalized-budget}
\end{equation}
Thus $a_b=0$ denotes identical projected states and $a_b=1$ denotes zero cross-covariance, rather than the largest mathematically possible error. Equation~(\ref{eq:normalized-budget}) assumes differentiable energies. With discrete common forcing kicks, it holds between kicks; a forced band's normalizer also has jumps that must be retained in its normalized balance. The primary SQG band is not directly forced. A stationary distribution permits $S_b(t)$ to fluctuate along each realization. For positive twin energies, $\rho_b=C_b/\sqrt{E_{t,b}E_{f,b}}$ satisfies $x_C=[2\sqrt{E_{t,b}E_{f,b}}/S_b]\rho_b$. A shell coordinate $1-\rho_k$ therefore coincides with normalized shell error for equal twin energies, and also at zero covariance. Away from zero covariance, equality requires equal twin energies. Negative correlations can give amplitudes above one; the approximation bounds below are explicitly restricted to intervals with amplitudes in $[0,1]$.

The error norm also matters when comparing scalar models. If $e=\sqrt{2\varepsilon}$ denotes the error norm (the root-mean-square error for a spatial-mean inner product), Eq.~(\ref{eq:logistic-reference}) becomes $\dot e=(\lambda_0/2)e[1-e^2/(2L)]$, a cubic saturation law in $e$. An amplitude-growth rate and an energy-growth rate differ by a factor of two in the infinitesimal exponential regime \citep{Bach2025}.

\subsection{A conditional invariant-ray proposition}
\label{sec:ray}
We now pass from individual realizations to ensemble-mean spectra. For twins with the same stationary marginal spectrum, let $U(k)$ denote the energy spectrum of either state and let $W(k,t)$ denote the real, symmetrized cross-spectrum, normalized so that their integrals are the single-state half-energy and cross-covariance, respectively. The radial wavenumber is $k$. Write $\Delta(k,t)=U(k)-W(k,t)$ for the decorrelated part. Then the ensemble-mean error is $\varepsilon(t)=2\int\Delta(k,t)\,\mathrm dk$ in the same half-energy convention. Integrals are over the participating spectral support; for discrete shells they are replaced by sums. Stationarity is an assumption for this reduction, not a statement that each realized shell energy is constant. Following the separation of correlated and decorrelated spectra in classical predictability theory \citep{Leith1971,LeithKraichnan1972}, we state the following assumed exchange class explicitly:
\begin{equation}
 \partial_t\Delta=\mathcal B[W,\Delta]+r_\Delta,
 \qquad W=U-\Delta,\qquad G(k)=\mathcal B[U,U](k).
 \label{eq:exchange-model}
\end{equation}
Here $\mathcal B$ is a specified, time-independent bilinear operator, linear in each spectral argument separately; it can couple different wavenumbers. The remainder $r_\Delta$ collects the contributions beyond that bilinear conversion, including unbalanced spectral transport, forcing/dissipation contributions, and memory effects. Equation~(\ref{eq:exchange-model}) defines a conditional model class. Identifying a particular closed kernel $\mathcal B$ and evaluating its remainder require additional closure information beyond the exact twin budget and the SQG calculations presented here.

A common spectral shape means $\Delta(k,t)=a(t)U(k)$: every participating scale has the same normalized decorrelation amplitude. We call the set of such spectra an invariant ray when initial data on it remain on it under the evolution. Only the segment $0\leq a\leq1$ is considered here. Assume $U>0$ on the participating support, $0<\int U\,\mathrm dk<\infty$, a well-posed evolution (existence and uniqueness for the initial data under consideration), and $r_\Delta=0$ whenever $\Delta=aU$ along the candidate ray. Substitution of $\Delta=a(t)U$ and $W=(1-a)U$ into Eq.~(\ref{eq:exchange-model}) gives
\begin{equation}
 U(k)\dot a=a(1-a)G(k),\qquad
 \frac{G(k)}{U(k)}=\lambda_*\ \Longrightarrow\ \dot a=\lambda_*a(1-a).
 \label{eq:ray-proposition}
\end{equation}
For a nonconstant interior ray trajectory, the ratio must be independent of $k$, since the same $\dot a/[a(1-a)]$ appears at every contributing scale. Conversely, if that ratio is constant and the remainder vanishes on the ray, the scalar logistic solution generates a solution of the spectral model, and uniqueness keeps initial data on the ray there. For $\lambda_*>0$ this is logistic growth, with $L=2\int U\,\mathrm dk$ and $\varepsilon=La$. The endpoints $a=0,1$ are trivial stationary solutions for every ratio $G/U$ under the stated remainder hypothesis.

Within the zero-remainder class, a nonzero constant $G/U$ is therefore necessary and sufficient for a nonconstant interior ray solution. A zero common ratio also preserves the ray, but every point on it is stationary. Vanishing remainder is a hypothesis of that equivalence, not a universal requirement for ray invariance or for logistic aggregate growth. For example, if $G/U=\lambda_*$ and $r_\Delta=\beta(a,t)U$, the remainder preserves the common spectral shape while the amplitude remains in the admissible interval, but the amplitude obeys
\begin{equation}
 \dot a=\lambda_*a(1-a)+\beta(a,t).
 \label{eq:parallel-remainder}
\end{equation}
This law is logistic with a specified constant rate $\mu$ only if $\beta=(\mu-\lambda_*)a(1-a)$ along the trajectory. A general parallel remainder changes the amplitude law. If $G/U$ varies across scales, a purely parallel remainder cannot cancel the part of the conversion tendency that changes the spectral shape. Matching $\lambda_*$ to an independently measured tangent energy rate is a further empirical test, distinct from deriving the logistic functional form.

\subsection{Identical logistic aggregates with different spectral dynamics}
\label{sec:two-shell-example}
\begin{figure}[t]
\centering
\includegraphics[width=0.96\textwidth]{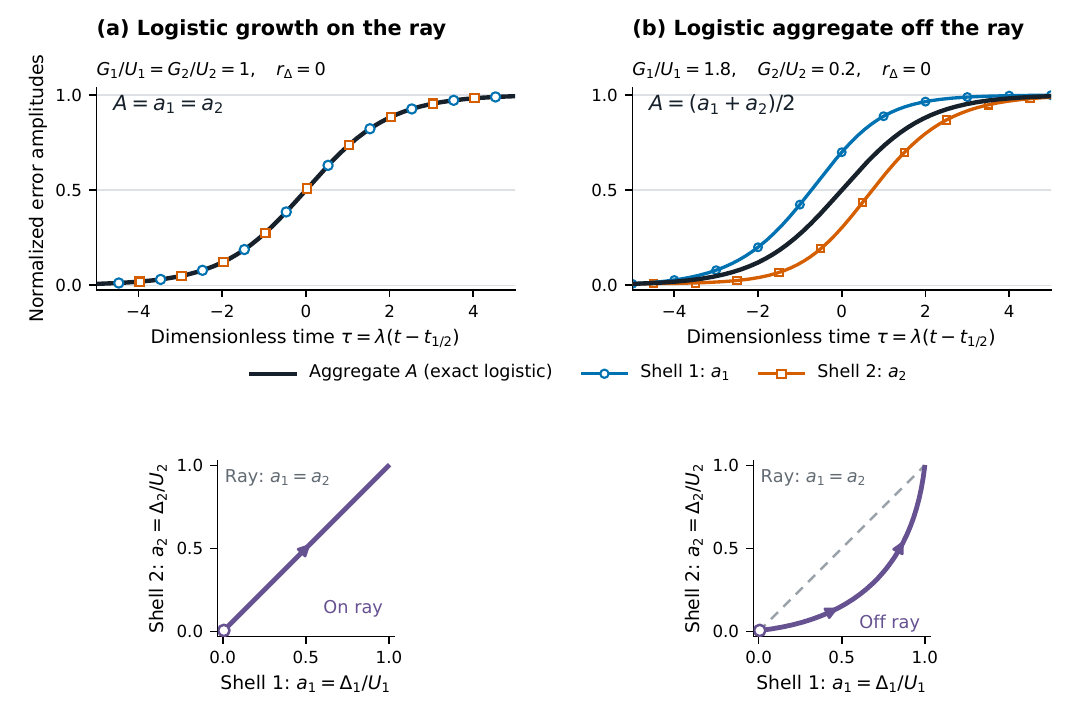}
\caption{Identical logistic aggregate growth with different spectral dynamics in the exact two-shell model of Eq.~(\ref{eq:two-shell-operator}). Both cases have $U_1=U_2=1$, $r_\Delta=0$, $\lambda=1$, and the same initial amplitude $A_0=(1+e^5)^{-1}$. The black curve is $A=(a_1+a_2)/2=\varepsilon/L$. (a) Equal conversion ratios, $G_1/U_1=G_2/U_2=1$, preserve the common-shape ray, and both shell amplitudes coincide with $A$. (b) Unequal ratios, $G_1/U_1=1.8$ and $G_2/U_2=0.2$, separate the shell amplitudes while leaving their mean exactly logistic. Lower plots show the spectral trajectories; arrows indicate increasing time, open circles mark the initial state, and the diagonal is $a_1=a_2$. These are analytical constructions, not SQG simulation results. Identical logistic aggregates therefore do not identify invariant-ray spectral dynamics.}
\label{fig:ray-aggregate}
\end{figure}

Figure~\ref{fig:ray-aggregate} illustrates both the ray construction and the failure of its converse for aggregate growth. We use an explicitly constructed model because the SQG diagnostics below do not establish the common-ratio and zero-remainder conditions for any archived trajectory. This example belongs to the bilinear class in Eq.~(\ref{eq:exchange-model}); its operator is not derived from SQG.

Take two discrete shells with fixed reference energies $U_1=U_2=1$ in chosen energy units. Let $a_i=\Delta_i/U_i$, $W_i=1-a_i$, $\lambda>0$, and $0\leq c\leq1$. Prescribe the autonomous bilinear operator
\begin{align}
 \mathcal B_1[W,\Delta]&=\frac{\lambda}{4}(\Delta_1+\Delta_2)
 [(3+c)W_1+(c-1)W_2],\nonumber\\
 \mathcal B_2[W,\Delta]&=\frac{\lambda}{4}(\Delta_1+\Delta_2)
 [-(1+c)W_1+(3-c)W_2],
 \label{eq:two-shell-operator}
\end{align}
with $\dot\Delta=\mathcal B[U-\Delta,\Delta]$ and $r_\Delta=0$. Define the aggregate amplitude and half-difference by $A=(a_1+a_2)/2$ and $d=(a_1-a_2)/2$. Adding and subtracting the component equations gives
\begin{equation}
 \dot A=\lambda A(1-A),\qquad
 \dot d=\lambda A[c(1-A)-2d].
 \label{eq:two-shell-reduction}
\end{equation}
The aggregate equation is logistic for every initial state in the unit square, not a rate fitted to one trajectory. In the energy convention used here, $\varepsilon=2(\Delta_1+\Delta_2)=4A$ and $L=4$, so $A=\varepsilon/L$. The polynomial vector field points inward or tangentially on each face of $0\leq a_1,a_2\leq1$ when $0\leq c\leq1$. This square is therefore invariant, and smoothness gives a unique global forward solution within it.

Evaluation on the reference spectrum gives $G/U=\lambda(1+c,1-c)$. For $c=0$ and initially equal amplitudes, $d$ remains zero and the spectrum stays on the ray. For $c>0$, an initial interior point on the ray has $\dot d=\lambda cA(1-A)>0$: the shell amplitudes immediately separate, although their mean follows the same logistic equation. The changing spectral contribution cancels in the sum by the prescribed operator. This establishes the logical nonconverse without modifying a remainder.

For the displayed solutions, let $\tau=\lambda(t-t_{1/2})$, where $t_{1/2}$ is the time at which $A=1/2$, and choose $A_0=(1+e^5)^{-1}$ and $d(-5)=0$. The exact trajectories are
\begin{equation}
 A(\tau)=\frac{1}{1+e^{-\tau}},\qquad
 d(\tau)=c\frac{[1-A(\tau)][A(\tau)-A_0]}{1-A_0},\qquad a_{1,2}=A\pm d.
 \label{eq:two-shell-solution}
\end{equation}
Panels~(a) and (b) use $c=0$ and $c=0.8$, respectively, with $\lambda=1$ and the same initial state. The common-shape ray is the diagonal in the lower plots. Thus an exact logistic aggregate can conceal an evolving distribution of error across scales. The construction demonstrates possibility within the stated model class; it does not establish that this cancellation explains the SQG curves.

\subsection{Averaging heterogeneous shell amplitudes}
\label{sec:aggregation}
On a fixed finite shell set, let normalized weights $w_k(t)\geq0$ satisfy $\sum_kw_k=1$, and write
\begin{equation}
 \dot a_k=\lambda_k(t)a_k(1-a_k)+q_k(t),\qquad A=\sum_kw_ka_k,\qquad
 \bar\lambda=\sum_kw_k\lambda_k.
 \label{eq:local-clock-model}
\end{equation}
Here $a_k$ is a normalized shell amplitude, $\lambda_k(t)$ is a specified shell growth coefficient (a local growth ``clock''), and $q_k(t)$ is the difference between the actual amplitude tendency and the proposed logistic term. Equation~(\ref{eq:local-clock-model}) is a local surrogate, meaning a proposed evolution law for each shell; it is distinct from the nonlocal exchange operator in Eq.~(\ref{eq:exchange-model}). Assume $a_k$ and $w_k$ are absolutely continuous and the displayed products are integrable. With weighted variance $V=\sum_kw_k(a_k-A)^2$ and $f_k=a_k(1-a_k)$, direct differentiation yields
\begin{equation}
 \dot A=\bar\lambda\{A(1-A)-V\}
       +\operatorname{Cov}_w(\lambda,f)
       +\sum_k\dot w_k(a_k-A)+\sum_kw_kq_k.
 \label{eq:weighted-identity}
\end{equation}
The weighted covariance is $\operatorname{Cov}_w(\lambda,f)=\sum_kw_k(\lambda_k-\bar\lambda)(f_k-\bar f)$, where $\bar f=\sum_kw_kf_k$. Thus $V$ measures variation in decorrelation amplitude, while the covariance measures whether shells with larger rates have larger logistic factors. The identity follows by writing $\sum_kw_k\lambda_k f_k=\bar\lambda\sum_kw_k f_k+\operatorname{Cov}_w(\lambda,f)$ and using $\sum_kw_k f_k=A(1-A)-V$. Normalization gives $\sum_k\dot w_k=0$, which permits the centered weight term. Fixed weights and zero residual recover the familiar variance--covariance decomposition. For an exact band-energy coordinate, choose $a_k=\varepsilon_k/S_k$ and $w_k=S_k/S_b$ on a fixed shell partition, where $S_k=E_{t,k}+E_{f,k}>0$ and $S_b=\sum_kS_k$. Then $A=\varepsilon_b/S_b$, and the weight term explicitly accounts for redistribution of the normalization across shells. This choice differs from the correlation amplitudes and truth-energy weights used in the shape diagnostic of Sect.~\ref{sec:spectral_diagnostics}.

Defining $\lambda_k=\dot a_k/[a_k(1-a_k)]$ from the same data makes $q_k=0$ by construction on interior samples. The resulting equality is descriptive; testing a local logistic mechanism requires independently specified rates or residuals. With instantaneous weights, $\dot A$ equals $\sum_kw_k\dot a_k$ plus the weight-evolution term; equality with the weighted shell derivatives holds when that term vanishes. A changing binary eligibility mask can introduce jumps and requires explicit boundary/jump accounting to extend the absolutely continuous fixed-domain statement. These distinctions apply directly to the exploratory calculation of Sect.~\ref{sec:results}.

\subsection{A finite-time approximation bound}
\label{sec:finite-time-bound}
Let $z$ solve $\dot z=\lambda_0z(1-z)$ for a fixed reference rate $\lambda_0\geq0$, and assume $a_k,A,z\in[0,1]$ on $[t_0,T]$. Define the defect $D(t)$ as the difference between the aggregate tendency and the reference logistic tendency. Its exact decomposition is
\begin{align}
 D(t)={}&(\bar\lambda-\lambda_0)A(1-A)-\bar\lambda V
       +\operatorname{Cov}_w(\lambda,f)\nonumber\\
       &+\sum_k\dot w_k(a_k-A)+\sum_kw_kq_k,\qquad
 \dot A=\lambda_0A(1-A)+D.
 \label{eq:logistic-defect}
\end{align}
If $D$ is integrable, subtraction of the reference equation and the bound $|x(1-x)-y(1-y)|\leq|x-y|$ for $x,y\in[0,1]$ give, by the integral Gronwall inequality,
\begin{equation}
 |A(t)-z(t)|\leq e^{\lambda_0(t-t_0)}|A(t_0)-z(t_0)|
          +\int_{t_0}^{t}e^{\lambda_0(t-s)}|D(s)|\,\mathrm ds.
 \label{eq:finite-time-bound}
\end{equation}
Thus the full defect controls a sufficient finite-time approximation bound, with shape variance contributing one of its terms. If $|D|\leq d_*$, the integral is at most $d_*[e^{\lambda_0(t-t_0)}-1]/\lambda_0$ for $\lambda_0>0$, and $d_*(t-t_0)$ for $\lambda_0=0$.

Define the weighted rate variance $\sigma_{\lambda,w}^2=\sum_kw_k(\lambda_k-\bar\lambda)^2$. For a bound based on absolute magnitudes, Cauchy--Schwarz gives $|\operatorname{Cov}_w(\lambda,f)|\leq\sigma_{\lambda,w}\sqrt{V}$, because $f(a)=a(1-a)$ is 1-Lipschitz on $[0,1]$. Consequently $|D|$ is bounded by $|\bar\lambda-\lambda_0|A(1-A)+|\bar\lambda|V+\sigma_{\lambda,w}\sqrt V+\sum_k|\dot w_k||a_k-A|+\sum_kw_k|q_k|$. This estimate is conservative and can be loose over long intervals. Applying it to band error additionally requires a consistent energy coordinate and the normalization terms in Eq.~(\ref{eq:normalized-budget}). The archived-data audit evaluates related instantaneous terms and an auxiliary bound for a sampled interpolation (Sect.~\ref{sec:physical-audit} and Appendix~\ref{app:evidence}); it does not certify this bound for the continuous physical trajectory.
\section{SQG experiments and diagnostics}
\label{sec:methods}
\subsection{Model, forcing, and sampling}
\label{sec:sqg_setup}
We analyze an existing forced surface-quasigeostrophic (SQG) ensemble. SQG represents advection of an active surface scalar by a horizontally nondivergent velocity determined from that scalar \citep{Held1995}. In the simulation convention,
\begin{equation}
 \partial_t\theta+J(\psi,\theta)=\nu\nabla^2\theta-\mu(-\nabla^2)^{-1}\theta+f,
 \qquad \widehat\psi(\boldsymbol{k})=\widehat\theta(\boldsymbol{k})/k,
 \label{eq:sqg_model}
\end{equation}
Here $\theta$ is the active surface scalar, $\psi$ is the streamfunction, $\nu$ is the small-scale diffusivity, and $\mu$ sets the large-scale damping. Hats denote Fourier coefficients. For $k=|\boldsymbol{k}|>0$, the velocity is $\boldsymbol v=(-\partial_y\psi,\partial_x\psi)$, and $J(\psi,\theta)=\boldsymbol v\cdot\nabla\theta$. The domain is a doubly periodic square of side $2\pi$; times are in model units. The energy is scalar half-variance, $E=\langle\theta^2\rangle/2$, and error energy is $\varepsilon=\langle(\theta_f-\theta_t)^2\rangle/2$, with spatial-mean angle brackets. The Fourier normalization gives $E=\sum|\widehat\theta|^2/(2N^4)$.

The solver uses a pseudospectral discretization on an $N\times N$ grid with square support $|k_x|,|k_y|\leq\lfloor N/3\rfloor$. Time integration uses fourth-order Runge--Kutta with integrating factors for the damping rate $\nu k^2+\mu/k^2$. At $N=256$, the viscosity is $\nu=0.0134$ and the time step is $\Delta t=0.0004$. At $N=512$, these values are $0.0067$ and $0.0002$, respectively. Both configurations use $\mu=1$. With the Reynolds number defined as $\mathrm{Re}=\epsilon_I^{1/3}(2\pi/k_f)^{4/3}/\nu$, the two configurations give $\mathrm{Re}=469.7$ and $939.3$. Resolution and viscosity change together, so the comparison represents different physical regimes. A fixed-parameter convergence sequence requires a separate design.

Forcing consists of fixed-modulus Fourier coefficients with random phases and Hermitian symmetry, applied as additive random increments (kicks) proportional to $\sqrt{\Delta t}$ after each deterministic step. The nominal band $k_f=3.5$, with width $0.5$, contains eight lattice modes at $k=\sqrt{13}$. Their modulus sets the expected scalar-energy injection to $\epsilon_I=24$. Each truth and its two forecasts receive the same forcing realization; different truth groups use distinct forcing seeds. Shells are unit-width radial bins centered on integer wavenumbers. The primary analysis band comprises shells 10--20, or $9.5\leq k<20.5$, disjoint from forcing.

Within each regime, four initial truth states were taken from one long reference integration (the nature run) at nature-run times 100, 115, 130, and 145, with the first state sampled at the end of a 100-unit spin-up. Each initializes two forecasts with separately seeded Gaussian perturbations in the primary band, scaled to initial error energy $10^{-6}(2E_{t,b})$. Forecasts extend to 120 units. The stationarity screen requires the absolute fitted change in total truth energy over $60\leq t\leq120$, divided by its mean on that interval, to be at most $0.15$. All eight truth groups pass. The screen assesses total-energy drift over this finite window; shell-energy variability and dependence between the separated initial states remain separate questions.

\subsection{Independent tangent clock and production}
\label{sec:tangent_clock}
A separately seeded tangent perturbation evolves along each truth under the tangent-linear SQG equation. It starts in the primary band with full-field norm $10^{-6}$ and is renormalized to that norm every $h=0.1$ unit. The band rate measured before each renormalization is
\begin{equation}
 \lambda_b(t_j)=h^{-1}\log[E_{\eta,b}(t_j^-)/E_{\eta,b}(t_{j-1}^+)],
 \label{eq:tangent_rate}
\end{equation}
where $\eta$ is the tangent perturbation and the superscripts $-$ and $+$ denote immediately before and after renormalization, respectively, and $E_{\eta,b}=\|\Pi_b\eta\|^2/2$. We use the arithmetic mean for $t_j\geq5$ as each truth group's $\lambda_b$. The two forecasts share this energy-growth rate, measured independently of their finite-error curves. It supplies the tangent clock, meaning the time coordinate rescaled by this rate.

Scalar diagnostics were stored every $0.02$ unit through $t=12$, every $0.25$ unit through $t=20$, and every unit thereafter, giving 733 records. The 125 field snapshots cover $t=0,0.1,\ldots,12$ and $t=20,40,80,120$. The projected production components are defined in Eq.~(\ref{eq:band-budget}). The plotted response and covariance coordinate are
\begin{equation}
 y=\frac{P_{\mathrm{linear}}+P_{\mathrm{TL,nl}}}{\lambda_b\varepsilon_b},
 \qquad x_C=\frac{2C_b}{E_{t,b}+E_{f,b}}.
 \label{eq:methods_response}
\end{equation}
The candidate relation $y=x_C$ concerns the linearized production; the full band-energy tendency also contains pure-error transfer and any numerical residual. The normalized amplitude additionally contains the $\dot S_b$ term.

\subsection{Spectral shape, effective clocks, and uncertainty}
\label{sec:spectral_diagnostics}
Shell energies and covariance define $\rho_k=C_k/(E_{t,k}E_{f,k})^{1/2}$ and $a_k^{(\rho)}=1-\rho_k$. The superscript distinguishes this correlation coordinate from the energy coordinate $\varepsilon_k/S_k$ in Sect.~\ref{sec:aggregation}. The prespecified shape diagnostic uses instantaneous truth-energy weights, renormalized over retained shells:
\begin{equation}
 R(t)=\frac{\sum_kw_k(a_k^{(\rho)}-\bar a)^2}{\bar a(1-\bar a)},\qquad
 \bar a=\sum_kw_ka_k^{(\rho)},\qquad w_k\propto E_{t,k}.
 \label{eq:shape_statistic}
\end{equation}
It uses unsmoothed amplitudes at field times $t\leq12$ with band correlation $\rho_b=C_b/(E_{t,b}E_{f,b})^{1/2}$ satisfying $0.15\leq\rho_b\leq0.85$. Shells contributing less than $10^{-3}$ of band truth energy are excluded. The shape statistic is retained only for a finite denominator with $0<\bar a<1$. At least two shells and five eligible times are required for a trajectory summary.

Effective finite-error rates define the clock-dispersion diagnostic:
\begin{equation}
 \lambda_{\mathrm{eff},k}=\frac{D_t\widetilde a_k^{(\rho)}}{a_k^{(\rho)}[1-a_k^{(\rho)}]},\qquad
 \Lambda_{\mathrm{disp}}=\mathrm{sd}_w[\log\lambda_{\mathrm{eff},k}],
 \label{eq:effective_shell_clock}
\end{equation}
Here $\mathrm{sd}_w$ is the weighted standard deviation of the natural logarithm, with the weights normalized over the eligible shells. A common change of rate units leaves this dispersion unchanged. The quantity $\widetilde a_k^{(\rho)}$ is a three-time moving average, and $D_t$ is the centered difference on the $0.1$-unit cadence. This implementation uses unsmoothed amplitudes in the denominator and endpoint mask. Only positive finite rates with $a_k^{(\rho)}\in[0.15,0.85]$ and sufficient shell energy enter the logarithm; excluded pairs are counted. These rates are inferred from the finite-error amplitudes. They measure neither $G/U$ nor the remainder $r_\Delta$ in the exchange model and differ from the separately integrated tangent rate. The prespecified shape and rate tolerances are $R\leq0.05$ and $\Lambda_{\mathrm{disp}}\leq0.10$; $R>0.25$ defines a diagnostic region of large spectral contrast, called the sharp-front region in the original protocol. A front here means a transition in decorrelation across wavenumbers; the threshold alone establishes neither its profile nor its propagation. In the separate exploratory local-clock calculation, both the derivative and the amplitude in its denominator use the smoothed series, and its retained shells must lie in the interior interval after smoothing.

Each trajectory is summarized by its time median. The two member summaries are averaged within each truth group, and the regime statistic is the median of four group values. The 95\,\% percentile intervals use 10\,000 bootstrap resamples of these groups with seed 20260811. Resampling preserves the grouping of times and members within each truth state. Interpretation remains limited by the small group count and unverified independence of initial truth states.

\subsection{Descriptive logistic comparison}
\label{sec:descriptive_comparison}
An exploratory two-means clustering partitions the trajectories into two groups using only the standardized trajectory summaries $(R,\Lambda_{\mathrm{disp}})$, after centering each diagnostic and dividing by its across-trajectory standard deviation. For each trajectory, $a_b=1-x_C$ is compared with $g(\tau)=(1+e^{-\tau})^{-1}$, where $\tau=\lambda_b(t-t_{1/2})$ and $t_{1/2}$ is interpolated from the first upward crossing of $a_b=1/2$. Thus the origin is retrospective. Over finite scalar samples with positive error energy and $0.15<\rho_b<0.85$, the score is
\begin{equation}
 \mathrm{RMSE}_i=\left\{n_i^{-1}\sum_{j=1}^{n_i}[a_{b,i}(t_j)-g(\tau_{ij})]^2\right\}^{1/2}.
 \label{eq:individual_rmse}
\end{equation}
Here $i$ identifies a trajectory, $j$ its eligible scalar records, and $n_i$ their count. The group scores average these individual errors. Binned production curves use $x_C$ bins after selection by $\rho_b$; their error bars are descriptive standard errors across trajectory-bin means, and they treat trajectories sharing a truth state as separate contributions.

\subsection{Relational null tests}
\label{sec:relational_nulls}
A relational null test changes the association between otherwise retained data to assess a particular coupling. Here the analysis tests the positive weighted regression slope of $y$ against $x_C$, allowing an intercept and assigning equal weight to each truth group and member. Its statistic, masks, and surrogate rules were fixed before the null tests were evaluated. The constructions are 9999 within-trajectory permutations of five-sample response blocks; all $4!\,2^4=384$ truth-response reassignments and member swaps at matching times; and 999 common Hermitian phase scrambles of the truth and its forecasts at each stored field time. Phase scrambling preserves modal amplitudes and cross-spectra while changing nonlinear triad production, which is recomputed from the fields. Predictors and recipient masks are retained in response reassignments. Upper-tail comparisons count ties against the observed statistic. The intersection--union rule treats the joint claim as supported only when every component null is rejected: all three tests must reject at $0.05$ in the primary $N=512$ regime. The evidence is conditional on these surrogate constructions, including truth-label exchangeability, and on the decision rule fixed beforehand.
\subsection{Audit of the existing fields and scalar records}
\label{sec:existing_audit}
A subsequent audit reconstructed moments and instantaneous spectral operators from the stored fields, checked the allocated energy increments in five disjoint bands spanning the retained modes and in the full retained spectrum, and evaluated normalization and fixed-shell averaging terms. It used the existing eight truth groups without further integrations or resampling. The primary-band dependencies pass; the nonprimary accounting failures and their interpretation are retained in Appendix~\ref{app:evidence}. On the fixed shells 10--20, let $P_b=P_{\mathrm{linear}}+P_{\mathrm{TL,nl}}+P_{\mathrm{pure}}$. Evaluating the operators on each archived state gives
\begin{equation}
 T_b^{\mathrm{op}}=-a_b\frac{\dot S_b^{\mathrm{op}}}{S_b},\qquad
 \dot a_b^{\mathrm{op}}=\frac{P_b}{S_b}+T_b^{\mathrm{op}},\qquad
 D_b^{\mathrm{op}}=\dot a_b^{\mathrm{op}}-\lambda_b a_b(1-a_b).
 \label{eq:audit_operator_defect}
\end{equation}
Here $a_b=\varepsilon_b/S_b$ and $\lambda_b$ is the independent band tangent rate. Superscript $\mathrm{op}$ denotes an operator evaluation on a saved state, rather than a finite-difference derivative of the sampled path. The primary band receives no direct forcing kick. A descriptive comparison specified after inspecting the audit matches the original lifecycle window: field times $t\leq12$ with $0.15<\rho_b<0.85$, retaining every shell in the fixed band. RMS denotes the square root of the arithmetic mean of the squared values. All RMS quantities in this comparison use these same times, with no temporal interpolation. For each member we define
\begin{equation}
 \mathcal C_b=\max\{\mathrm{RMS}(\dot a_b^{\mathrm{op}}),\,
                 \mathrm{RMS}[\lambda_b a_b(1-a_b)]\}.
 \label{eq:audit_characteristic}
\end{equation}
Member ratios are averaged within each truth group and the four group values summarized by their median. The separate normalizer coefficient of variation, $\mathrm{sd}(S_b)/\mathrm{mean}(S_b)$, weights all 733 scalar records equally; it is not a uniform-time statistic.

\section{Finite-band evidence from the SQG experiments}
\label{sec:results}
\subsection{Spectral coherence and effective-rate heterogeneity}
The sampled band contains a resolved decorrelation transition and a spectrum steeper than the canonical $k^{-5/3}$ constant-flux SQG variance-cascade scaling. Log--log fits of the truth shell spectrum over $k=10$--$20$ give median slopes $n=3.24$ at $N=256$ and $n=2.70$ at $N=512$; the respective ranges across truth-state medians are $[3.18,3.26]$ and $[2.63,2.72]$. These values characterize the finite sampled interval. Establishing a front reduction additionally requires a documented inertial range, a moving-front profile, and an independent front-speed measurement. The steeper regime also lies outside the scale-local strain interpretation used for the canonical front examples in Appendix~\ref{app:front}.

Table~\ref{tab:diagnostics} reports the shape and effective-rate summaries. The shape statistic lies near the prespecified $0.05$ threshold and varies among truth states. Effective-rate dispersion is substantially above its $0.10$ threshold in both regimes. All 16 trajectories exceed the latter threshold. Thus no trajectory meets both empirical screening criteria for a common-amplitude reduction. These screens assess similarity in shape and inferred rate; they are not direct tests of the operator condition $G/U=\lambda_*$ or of a vanishing spectral remainder. This is a finite-band diagnostic finding; inferred effective rates also contain the influence of transport, normalization, and unresolved terms.

\begin{table}[t]
\caption{Prespecified finite-band diagnostics. Intervals are 95\,\% truth-group bootstrap intervals. Four temporally separated truth states contribute to each regime, with two perturbations per state; these intervals summarize the sampled truth groups, with state independence and resolution convergence requiring separate evidence.}
\label{tab:diagnostics}
\centering
\begin{tabular}{lcccc}
\tophline
Regime & Median $R$ & Interval & Median $\Lambda_{\mathrm{disp}}$ & Interval\\
\middlehline
$N=256$ & 0.0433 & [0.0350, 0.0567] & 0.3741 & [0.2136, 0.4890]\\
$N=512$ & 0.0619 & [0.0481, 0.0649] & 0.3555 & [0.2764, 0.4297]\\
\bottomhline
\end{tabular}
\end{table}

\subsection{Retrospective agreement with the logistic reference}
\begin{figure}[t]
\centering
\includegraphics[width=0.96\textwidth]{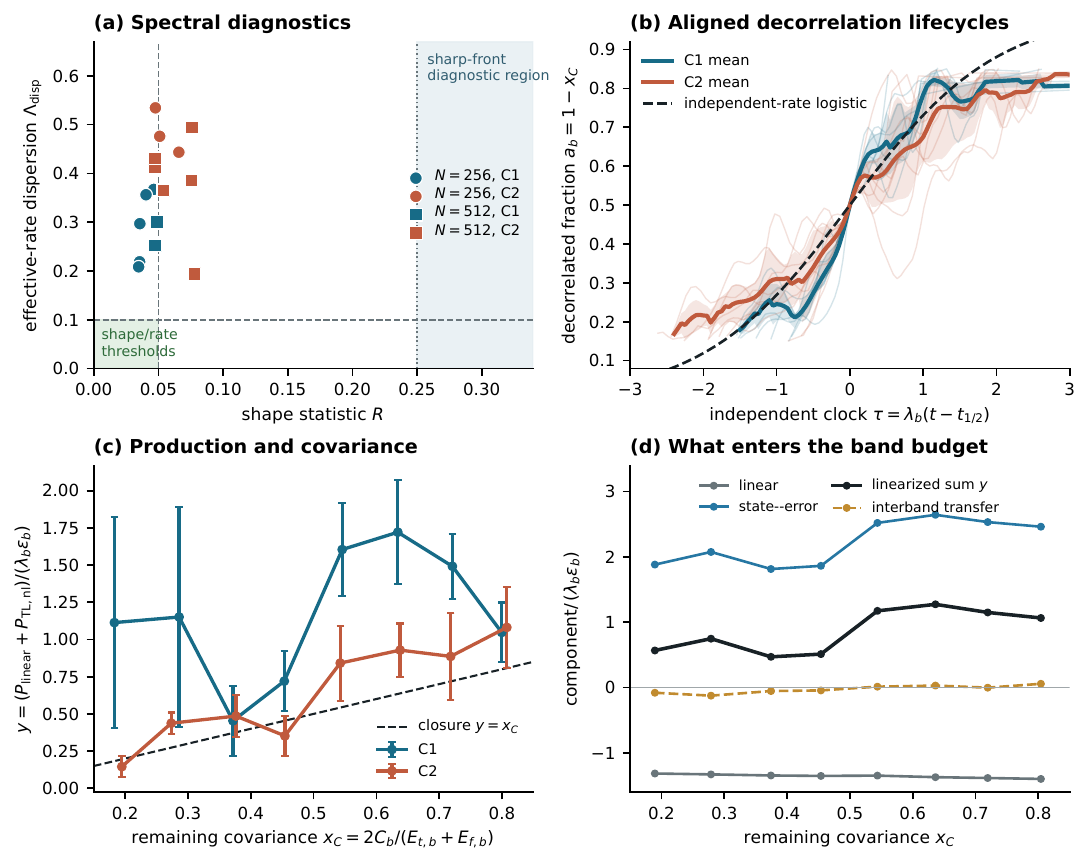}
\caption{Finite-band SQG diagnostics. (a) Shape statistic versus effective-rate dispersion for 16 trajectories. Symbols identify regimes; colors identify the descriptive two-means groups. The small shaded box marks the joint shape/rate thresholds only; the right-hand region is a prespecified sharp-front diagnostic threshold. Physical residuals require separate control in both regions. (b) Individual and mean decorrelation fractions versus independent tangent clock, aligned by each observed halfway crossing; shading shows the interpolated 16th--84th percentile range where at least two trajectories contribute. (c) The normalized production response compared with $y=x_C$. Error bars are descriptive trajectory-based standard errors; they treat trajectories sharing a truth state as separate contributions. (d) Normalized budget components, with pure error self-interaction retained as interband transfer.}
\label{fig:lifecycle}
\end{figure}

The descriptive partition separates seven lower-dispersion trajectories from nine higher-dispersion trajectories (Fig.~\ref{fig:lifecycle}a). Their mean trajectory-level logistic root-mean-square deviations are 0.080 and 0.093. These values average individual trajectory errors, as defined in Eq.~(\ref{eq:individual_rmse}); errors of the two plotted mean curves are distinct quantities. Individual values span approximately 0.054--0.187. The two group averages are therefore comparable, even though the partition was formed from the spectral diagnostics alone, without reference to these deviations. Predictive equivalence and preference among competing growth models require additional tests.

Figure~\ref{fig:lifecycle} makes the distinction concrete: the retrospective lifecycle agreement in panel~(b) coexists with departures of the binned production from $y=x_C$ in panel~(c). This comparison is descriptive because the production error bars treat trajectories sharing a truth state as separate contributions. It illustrates the additional evidence required to connect curve agreement to a production mechanism.

The independent tangent rate constrains the time scale, but the observed halfway crossing supplies each time origin. Figure~\ref{fig:lifecycle}b is therefore a retrospective interior-lifecycle comparison. Its interpolated mean has a varying contributing sample, so its interpretation is restricted to the sampled interior lifecycle; testing the complete onset-to-saturation curve requires additional evidence. Every trajectory lies outside the prespecified $R>0.25$ sharp-front diagnostic region. A broad front can nevertheless appear coherent within a narrow band; classifying similarity profiles requires information beyond that threshold.

\subsection{Instantaneous decomposition and physical interpretation}
For the exploratory calculation, apply the fixed-weight part of Eq.~(\ref{eq:weighted-identity}) at each eligible time to the smoothed correlation amplitudes and inferred rates, using instantaneous truth-energy weights normalized over the retained shells. Denote their weighted mean by $A_s$, their variance by $V_s$, and the inferred weighted rate by $\bar\lambda_s$. With $f_{s,k}=\widetilde a_k^{(\rho)}[1-\widetilde a_k^{(\rho)}]$, the baseline is $B_s=\bar\lambda_s A_s(1-A_s)$. The reported fractions are
\begin{equation}
 F_{\mathrm{shape}}=-\frac{\bar\lambda_sV_s}{B_s},\qquad
 F_{\mathrm{cov}}=\frac{\operatorname{Cov}_w(\lambda_s,f_s)}{B_s},\qquad
 F_{\mathrm{net}}=F_{\mathrm{shape}}+F_{\mathrm{cov}}.
 \label{eq:exploratory-fractions}
\end{equation}
The subscript $s$ distinguishes this smoothed, sample-selected diagnostic from the energy-coordinate identity and the unsmoothed shape statistic $R$. Explicitly, $\lambda_{s,k}=D_t\widetilde a_k^{(\rho)}/f_{s,k}$ and $\bar\lambda_s=\sum_kw_k\lambda_{s,k}$. Finite signed rates are retained, without the positive-rate filter used for $\Lambda_{\mathrm{disp}}$, and samples require $|B_s|>10^{-12}$.

\begin{figure}[t]
\centering
\includegraphics[width=0.90\textwidth]{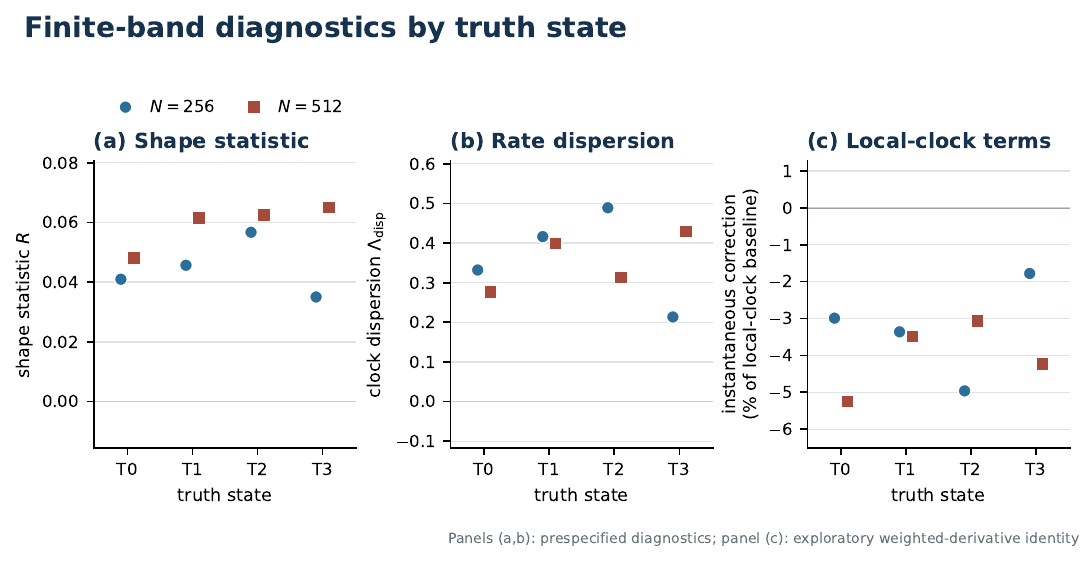}
\caption{Diagnostics by truth state. (a,b) Prespecified shape and effective-rate summaries. (c) The exploratory instantaneous local-clock correction, normalized by its inferred local-clock baseline. The total correction to band-energy dynamics also requires the remaining physical and normalization terms. T0--T3 denote the temporally separated truth states within each regime; the two perturbations share each state.}
\label{fig:truth-diagnostics}
\end{figure}

At $N=256$ and $N=512$ respectively, the exploratory local-clock calculation returns median shape fractions of $-2.16$ and $-3.04$\,\%, signed rate--shape fractions of $-0.95$ and $-0.41$\,\%, and net fractions of $-3.18$ and $-3.86$\,\%. These medians summarize each quantity separately, whereas additivity holds sample by sample before taking medians. Across truth groups, the net fractions range over $[-4.96,-1.78]$\,\% and $[-5.25,-3.07]$\,\% (Fig.~\ref{fig:truth-diagnostics}). The sample-wise algebraic residual is at most $8.9\times10^{-16}$.

These are fractions of the instantaneous baseline formed with the inferred weighted mean rate. They quantify a small correction to the weighted mean of shell derivatives on the selected samples. The numerical identity covers weighted shell derivatives; computing the tendency of their mean additionally requires weight evolution and changes in the eligible shell set. With unequal twin energies, conversion from shell correlations to normalized band energy also requires the energy-ratio factors. These percentages alone therefore do not evaluate the full defect in Eq.~(\ref{eq:logistic-defect}) or the finite-time bound. Section~\ref{sec:physical-audit} reports the subsequent audit in the consistent band-energy coordinate. The separately reported median shape and covariance contributions have the same sign and do not establish cancellation as an explanation of the band curve. The exact construction in Fig.~\ref{fig:ray-aggregate} shows that cancellation can preserve a logistic aggregate. It does not supply evidence that this mechanism accounts for the observed SQG trajectories.

The shape fractions in this exploratory calculation differ from the prespecified $R$ summary because smoothing and the per-shell interior mask select a different sample. The resulting values differ by roughly a factor of two. The analysis retains both definitions as distinct diagnostics with different sample selections.

\subsection{Measured normalization and physical tendency}
\label{sec:physical-audit}
The completed audit shows why the small exploratory correction does not establish a band-energy closure. Across the stored scalar records, the primary-band normalizer has coefficients of variation 0.2924 and 0.2488 at $N=256$ and $N=512$, respectively. These fluctuations are compatible with the separate late-time total-energy stationarity screen.

In the matched lifecycle window of Sect.~\ref{sec:existing_audit}, $\mathrm{RMS}(D_b^{\mathrm{op}})/\mathcal C_b$ is 0.8513 and 0.8668. The corresponding ratios for normalization and interband transfer are
\begin{align}
 \frac{\mathrm{RMS}(T_b^{\mathrm{op}})}{\mathrm{RMS}(P_b/S_b)}
   &=0.7345,\ 0.7295,\nonumber\\
 \frac{\mathrm{RMS}(P_{\mathrm{pure}}/S_b)}{\mathrm{RMS}(P_b/S_b)}
   &=0.3936,\ 0.3560,
 \label{eq:audit-physical-ratios}
\end{align}
in the same regime order. These last ratios compare RMS magnitudes with the RMS of the signed net production; they are not additive fractions or causal attributions. The window retains 5--18 field times per member at $N=256$ and 8--13 at $N=512$, with all four truth groups in each regime represented. The instantaneous discrepancy from the independently clocked logistic tendency therefore remains substantial on the sampled lifecycle. This descriptive calculation neither isolates a cause of that discrepancy nor rejects an ensemble-mean or stochastic logistic model. In particular, it does not identify the bilinear closure kernel or its remainder in Sect.~\ref{sec:ray}.

\subsection{The separate covariance-coupling test}
The relational-null analysis used the complete linearized production observable and the independently measured tangent rate. Its time-block and phase-scramble tests returned $p=0.011$ and $p=0.001$, whereas cross-truth reassignment returned $113/384$ exceedances, or $p=0.294$. The prespecified intersection--union rule required all three tests to reject at $0.05$. Because not all tests reject, the combined criterion is not satisfied. Truth-specific covariance coupling remains unresolved in this four-truth-state experiment; this is neither a demonstration of that coupling nor evidence for the local-clock decomposition as an alternative mechanism. The recorded values and numerical reconstruction checks are given in Appendix~\ref{app:evidence}.
\section{Implications for interpreting reduced forecast-error dynamics}
\label{sec:discussion}
\subsection{From a fitted curve to a testable approximation}
The practical question is what information is needed beyond the curve itself. An independent tangent rate is useful because a fitted finite-error rate can absorb effects of initial conditions, spectral redistribution, dissipation, and changing normalization. A spectral shape statistic supplies different information: it measures the departure from a shared amplitude across the selected scales. Control of the physical budget requires these diagnostics together with the remaining budget terms, and identifying a complete closure requires evidence across that full set.

The generalized averaging relation supplies a way to organize the missing information. Rate variation relative to the reference clock, amplitude variance, rate--shape covariance, evolving weights, and unresolved shell tendencies all contribute to the instantaneous defect. The finite-time bound translates an independently bounded defect into a bound on trajectory error over the specified interval. It is deliberately conservative, and can grow too rapidly to be useful over long intervals. Its role is to state an explicit sufficient condition for approximation accuracy based on the complete defect. The completed audit provides snapshot estimates of the physical defect and a bound for an auxiliary sampled path. Certification for the continuous physical trajectory still requires control between saved states, as explained in Appendix~\ref{app:evidence}.

Table~\ref{tab:evidence-map} summarizes the assumptions, available evidence, and remaining measurements for each level of interpretation.

\begin{table}[t]
\caption{Evidence map for interpreting a scalar forecast-error curve. Analytic statements apply under their listed assumptions; the diagnostics reported here provide the stated empirical evidence.}
\label{tab:evidence-map}
\centering
\small
\setlength{\tabcolsep}{4pt}
\begin{tabular}{>{\raggedright\arraybackslash}p{0.17\textwidth}>{\raggedright\arraybackslash}p{0.24\textwidth}>{\raggedright\arraybackslash}p{0.26\textwidth}>{\raggedright\arraybackslash}p{0.25\textwidth}}
\tophline
Statement & Assumptions & Available evidence & Remaining measurement\\
\middlehline
Exact error budget & Bilinear dynamics, a conservative norm, common forcing, and a fixed projection & Analytic balance; reconstructed moments, operators, and primary increments & Time-step and resolution convergence; control between stored states\\[4pt]
Conditional invariant ray & Stationary $U$, a common ratio $G/U$, and a remainder that vanishes on the ray & Conditional proof; exact two-shell nonconverse; empirical proxy diagnostics & A specified physical kernel and its remainder; comparison with an independent tangent rate\\[4pt]
Instantaneous averaging identity & Fixed shell set; absolutely continuous amplitudes and normalized weights & Exact identity; inferred-rate diagnostics; separate physical audit & Independent shell closure coefficients and continuous residual control\\[4pt]
Finite-time approximation bound & Fixed shell domain, absolutely continuous amplitudes and weights, amplitudes in $[0,1]$, and an integrable complete defect & Analytic Gronwall bound; auxiliary sampled-path calculation & Intersample control for a continuous-trajectory certificate\\[4pt]
Similarity-front law & A convergent similarity profile, a power-law spectrum, and a prescribed front speed & Conditional derivation and controlled formula checks & A similarity interval, independently measured slope and speed, and band-edge and physical-transfer terms\\
\bottomhline
\end{tabular}
\end{table}

A useful diagnostic sequence follows. First define the error norm, averaging operation, scale support, and reference energy. Then examine the spectral profile and the temporal variability of its weights. Compare effective-rate structure with an independently measured small-error clock, retaining the distinction between inferred rates and closure coefficients. Finally evaluate the complete physical tendency and normalization terms before assessing a quantitative scalar approximation. Applied to a compact ensemble representation, this sequence distinguishes preservation of the aggregate lifecycle from preservation of the scale distribution of uncertainty. The latter matters when different scales influence different forecast quantities \citep{UboldiTrevisan2015,ZagarEtAl2017}.

\subsection{Scope of the evidence and the next discriminating test}
The SQG example concerns perfect-model twins with common forcing and a particular finite band. Model error, observation error, and imperfect ensemble calibration require extensions of this experimental setting. These additional sources of uncertainty are central in operational forecasting and in alternative scalar descriptions \citep{BednarKantz2024,Bach2025}. The present framework complements those descriptions by making the spectral interpretation of a scalar law conditional and measurable.

Several limitations determine what a subsequent test must add. The two regimes change viscosity and resolution together; numerical convergence requires a separate fixed-parameter comparison. Four truth states per regime constrain inference, and their temporal separation specifies the sampling design, while statistical independence remains an additional assumption. Positive-rate and interior-amplitude masks limit the effective-rate diagnostic to a selected part of the lifecycle. The retrospective alignment removes onset-time variation. The audit finds substantial instantaneous discrepancies from the independent logistic reference, despite comparable retrospective curve scores. Inferred-rate identities alone cannot explain those discrepancies or supply a physical closure.

The completed audit supplies measurements that were absent from the initial exploratory decomposition. It confirms the need to retain the normalizer and physical transfer terms, and it records the numerical accounting failures without relaxing their tolerance. A subsequent convergence experiment should hold physical parameters fixed while varying numerical resolution and time step. A dedicated moving-front test would serve a different purpose: it must establish a similarity interval, measure spectral slope and front speed independently, and retain finite-band edge and transfer terms. Neither experiment is performed here.

\conclusions
A logistic forecast-error curve summarizes an aggregate multiscale lifecycle; identifying a common spectral growth mechanism requires additional spectral and budget evidence. Within the stated bilinear exchange class with a vanishing remainder on the ray, a nonzero common conversion coefficient is necessary and sufficient for a nonconstant interior common-shape solution; a positive coefficient gives logistic growth. The exact two-shell construction shows that logistic aggregate growth can also occur while that shape evolves. The general averaging relation explains which additional terms enter when these conditions are relaxed, and its finite-time bound states what must be measured to quantify the approximation.

The surface-quasigeostrophic experiments exhibit heterogeneous effective shell rates together with moderate retrospective agreement with an independently clocked logistic reference. Their exploratory decomposition is an instantaneous identity with inferred rates. The subsequent audit reproduces the primary-band summaries but finds substantial discrepancies from the independently clocked normalized logistic tendency and substantial normalization contributions, while preserving five failed highest-band increment checks. These findings constrain the interpretation of the scalar comparison without identifying its cause. The separate relational-null result also leaves truth-specific coupling unresolved. These limits specify a concrete route from descriptive forecast-error curves to testable reduced dynamics.

\appendix
\section{Conditional similarity-front reduction}
\label{app:front}
Classical two-field spectral predictability theory describes relative-error profiles that propagate toward lower wavenumbers \citep{LeithKraichnan1972,Ngan2009}. As an idealization of a finite similarity interval, extend the following power-law spectrum and error profile to $0<k<\infty$:
\begin{equation}
 U(k)=A_Uk^{-n},\qquad \Delta/U=r(k/k_e),\qquad
 \dot k_e=-\gamma k_e^{1+\beta},\qquad \gamma>0,
 \label{eq:front-profile}
\end{equation}
with $A_U>0$, $k_e>0$, $n>1$, $0\leq r\leq1$, and $0<I_n=\int_0^\infty x^{-n}r(x)\,\mathrm dx<\infty$. Here $k_e(t)$ locates the decorrelation front, $r$ is its fixed dimensionless profile, and $\beta$ specifies the wavenumber dependence of its propagation rate. The extension of $U$ is not an integrable whole-domain energy spectrum for $n>1$; only the decorrelated-spectrum integral is assumed finite. In particular, a fully decorrelated low-wavenumber end makes this integral diverge for $n>1$. Changing variable to $x=k/k_e$ gives $\varepsilon=B_nk_e^{1-n}$ with $B_n=2A_UI_n$. Differentiation and elimination of $k_e$ yield
\begin{equation}
 \dot\varepsilon=C_{n,r}\varepsilon^p,\qquad
 p=\frac{\beta+1-n}{1-n},\qquad
 C_{n,r}=(n-1)\gamma B_n^{1-p}.
 \label{eq:front-law}
\end{equation}
For $p\ne1$, $\varepsilon(t)^{1-p}=\varepsilon(t_0)^{1-p}+(1-p)C_{n,r}(t-t_0)$; for $p=1$ growth is exponential. These relations eliminate a specified front coordinate and inherit the front speed from the assumed dynamics.

If $U$ is a velocity-energy shell spectrum and the front's logarithmic propagation rate is proportional to the scale-local strain, the additional speed assumption gives $\beta=(3-n)/2$ and $p=(3n-5)/[2(n-1)]$ \citep{Rotunno2008}. The idealized exponents $n=5/3$, $2$, and $3$ then yield linear, square-root-tendency, and exponential error-energy growth, respectively. For $n=2$, the tendency is proportional to $\sqrt{\varepsilon}$, so $\sqrt{\varepsilon}$ grows linearly in time over the assumed similarity interval. Thus the square root describes the dependence of the tendency on error energy; the error energy itself is the square of a linear function of time. In the nondimensional SQG inversion used here, velocity energy and active-scalar variance have the same shell weighting, allowing that comparison \citep{Held1995}. For a barotropic-vorticity model the $k^{-3}$ example refers to kinetic energy; the corresponding vorticity-variance spectrum carries an additional factor $k^2$. The $n=3$ case is marginal for strain locality; logarithmic enstrophy-cascade corrections qualify the pure-power-law example \citep{Kraichnan1971}. Spectra steeper than $k^{-3}$ fall outside this local-strain interpretation. Bounded error energy remains compatible with transient superlinear growth: a finite similarity interval can terminate before a formal extrapolated divergence.

For a fixed band $[k_-,k_+]$, set $z_\pm=k_\pm/k_e$ and $I_n(z_-,z_+)=\int_{z_-}^{z_+}x^{-n}r(x)\,\mathrm dx$. For a sufficiently regular profile the exact derivative of its band integral is
\begin{align}
 \varepsilon_b&=2A_Uk_e^{1-n}I_n(z_-,z_+),\nonumber\\
 \dot\varepsilon_b&=2A_U\gamma k_e^{\beta+1-n}
 \{(n-1)I_n-z_-^{1-n}r(z_-)+z_+^{1-n}r(z_+)\}.
 \label{eq:front-band}
\end{align}
The endpoint terms result from differentiating the similarity profile over fixed physical limits. They are distinct from interband transfer in the dynamical budget. An empirical test must identify a similarity interval, specify the energy norm, measure $n$ and front speed independently, and evaluate both the endpoint terms and the full projected physical balance. Such a test requires spectral and dynamical evidence beyond the eleven-shell SQG diagnostics presented here.
\section{Evidence status and numerical verification}
\label{app:evidence}
The exact error budget, the conditional exchange proposition, and the generalized averaging identities are analytic statements. The finite-time inequality is likewise analytic; the audit evaluates it only for an auxiliary sampled path. The finite-band shape and effective-rate summaries use the diagnostic definitions fixed in advance. The two-means visualization and the stricter local-clock decomposition were carried out after the data had been examined and are therefore exploratory. All empirical flow integrations, parameter fits, resampling studies, and hypothesis tests reported here use the simulation set described in Sect.~\ref{sec:methods}. Figure~\ref{fig:ray-aggregate} instead evaluates the exact analytical solution in Eq.~(\ref{eq:two-shell-solution}); it uses no SQG samples, fitted rates, or new flow integration.

Reconstruction of truth energy, forecast energy, and covariance from the shell moments has relative errors $4.04$--$4.23\times10^{-8}$. These checks establish consistency of the stored moments and conventions. They are distinct from temporal discretization convergence or validation of a scalar closure. Controlled verification of the front formulas used $r(x)=x^m/(1+x^m)$, with the profile integral reproduced to relative error $5.86\times10^{-13}$ and the finite-band derivative to $1.30\times10^{-10}$. These are checks of the analytic formulas, not tests of the simulations.

The subsequent audit, whose records are included in the deposited materials, assessed allocated energy increments in five disjoint diagnostic bands and the full retained spectrum. We report three of these bands here: the primary band $9.5\leq k<20.5$, an intervening band
$20.5\leq k<32.5$, and the highest band $32.5\leq k\leq\sqrt2\lfloor N/3\rfloor$, intersected with the square dealiasing support of Sect.~\ref{sec:sqg_setup}. The upper radii are $85\sqrt2$ for $N=256$ and $170\sqrt2$ for $N=512$. For each band the audit compares the sum of the
four archived, step-accumulated contributions---linear, tangent-linear nonlinear, pure-error nonlinear, and
numerical residual---with the observed increment,
\begin{equation}
 r_b=\frac{\mathrm{RMS}\big(\sum_{\alpha=1}^{4}\Delta\varepsilon_b^{(\alpha)}
 -\Delta\varepsilon_b^{\mathrm{obs}}\big)}{\mathrm{RMS}\big(\Delta\varepsilon_b^{\mathrm{obs}}\big)},
 \label{eq:increment-audit}
\end{equation}
each value summarizing 732 stored output intervals, against a tolerance of $10^{-4}$ fixed in
advance. In the primary band all 16 trajectories pass, with $r_b$ between $9.36\times10^{-6}$ and
$1.385\times10^{-5}$. The intervening band also passes, with $r_b$ between $2.8215\times10^{-5}$
and $4.0126\times10^{-5}$ at $N=256$ and between $3.2592\times10^{-5}$ and $3.6936\times10^{-5}$
at $N=512$. In the highest band five of the eight $N=512$ trajectories exceed the tolerance, with
$r_b$ between $1.0161\times10^{-4}$ and $1.1264\times10^{-4}$. All eight highest-band checks at $N=256$, the other three at $N=512$, and all full-support increment checks pass. The largest discrepancies among these reported bands occur in the highest band; the smallest occur in the primary band. These are
increment-accounting discrepancies, not fractions of trajectory error. The highest band shares no
modes with shells 10--20, and the primary-band diagnostics do not take the highest-band allocated
increments as inputs; the field-to-moment and instantaneous-operator reconstruction checks pass in
every band, and the summaries reported in Sect.~\ref{sec:results} reproduce. The primary-band
result is therefore a statement about the internal consistency of the primary-band diagnostics. It
is not a convergence result: the bands remain dynamically coupled through nonlinear interactions,
so disjoint spectral support does not establish that the primary band is insensitive to numerical
error elsewhere, and neither temporal nor spatial convergence is tested here.

\begin{table}[ht]
\caption{Relational-null results and numerical checks. The cross-truth group includes all 384 allowed truth/member reassignments, with ties counted against the observed statistic. The cross-truth reassignment result leaves the required conjunction of the three null tests unmet.}
\label{tab:nulls}
\centering
\begin{tabular}{lcc}
\tophline
Check & Recorded value & Threshold set in advance\\
\middlehline
Time-block null & $p=0.011$ & $p\leq0.05$\\
Cross-truth reassignment & $p=113/384=0.294$ (rounded) & $p\leq0.05$\\
Phase-scramble null & $p=0.001$ & $p\leq0.05$\\
Field recomputation relative error & $4.5\times10^{-7}$ & $\leq10^{-4}$\\
Phase-invariance error & $3.0\times10^{-9}$ & $\leq10^{-5}$\\
\bottomhline
\end{tabular}
\end{table}

The audit also evaluated the finite-time bound for a piecewise-linear path through consecutive 0.1-unit samples. This is an auxiliary sampled-path calculation: each eligible segment restarts its reference, and quadrature agreement supplies a numerical convergence check rather than a certified enclosure. A bound for the continuous physical trajectory or an independent closure prediction remains unavailable because the intersample error envelope and independently constrained closure residuals have not been established.

The exact truth-state sampling times and the construction of trajectory-level error summaries are documented in Sect.~\ref{sec:methods}.
\noappendix

\codedataavailability{The archived SQG analysis code, comprising the diagnostic definitions, the empirical figure scripts, and the configuration and protocol files, is openly available under the MIT license at \url{https://github.com/mqpena/logistic-error-growth-limits}. The simulation data and reproduction materials, comprising the saved model fields, the initial conditions, the stationarity diagnostics, the compact trajectory vectors, and the diagnostic records, are deposited, with data under a Creative Commons Attribution 4.0 license and original code under the MIT license, at \url{https://doi.org/10.5281/zenodo.22822514}.}

\authorcontribution{Malaquias Pe\~na developed the study and is the sole manuscript author.}
\competinginterests{The author declares that there are no competing interests.}
\begin{acknowledgements}
The author acknowledges support from NOAA-OAR-JTTI under award NA22OAR4590179 and is grateful to Zoltan Toth and the late Eugenia Kalnay for discussions that helped motivate this research. The author received AI assistance with figure-script preparation and computational audits. The author is responsible for the scientific content and the manuscript.
\end{acknowledgements}

\end{document}